\documentclass[aps,prl,twocolumn,floatfix,showpacs,hyperref,amsmath,amssymb,nofootinbib]{revtex4}

\usepackage{mathtools}
\usepackage{bbm}
\usepackage{color}
\usepackage{graphicx}
\usepackage{hyperref}
\usepackage{IEEEtrantools}
\usepackage{txfonts}

\DeclareFontFamily{OT1}{pzc}{}
\DeclareFontShape{OT1}{pzc}{m}{it}{<-> s * [1.10] pzcmi7t}{}
\DeclareMathAlphabet{\mathpzc}{OT1}{pzc}{m}{it}

\newcommand{\eq}[1]{\begin{equation}#1\end{equation}}
\newcommand{\eqa}[1]{\begin{IEEEeqnarray}{rCl}#1\end{IEEEeqnarray}}

\newcommand{\hsp}{\hspace{1pt}}

\begin{document}

\title{Dynamics of dissipative Bose-Einstein condensation}

\author{S.\ Caspar}
\author{F.\ Hebenstreit}
\author{D.\ Mesterh\'{a}zy}
\author{U.-J.\ Wiese}
\affiliation{Albert Einstein Center for Fundamental Physics, Institute for Theoretical Physics, University of Bern, 3012 Bern, Switzerland}

\begin{abstract}
  We resolve the real-time dynamics of a purely dissipative $s=1/2$ quantum spin or, equivalently, hard-core boson model on a hypercubic $d$-dimensional lattice. 
  The considered quantum dissipative process drives the system to a totally symmetric macroscopic superposition in each of the $S^3$ sectors.
  Different characteristic time scales are identified for the dynamics and we determine their finite-size scaling.
  We introduce the concept of cumulative entanglement distribution to quantify multiparticle entanglement and show that the considered protocol serves as an efficient method to prepare a macroscopically entangled Bose-Einstein condensate.
\end{abstract}

\pacs{75.10.Jm, 03.75.Gg, 67.85.De, 03.67.Bg}

\maketitle

{\it Introduction.} 
After the ground-breaking discovery of Bose-Einstein condensation in ultracold quantum gases \cite{Anderson:1995gf,Davis:1995pg}, the cooling of clouds of atoms to nanokelvin temperatures has become a daily routine in atomic physics laboratories worldwide. 
However, our understanding of the underlying out-of-equilibrium process from first principles is far from complete. 
The reason for this is twofold. 
Conventional Monte Carlo methods are typically not applicable to the problem of real-time dynamics of quantum many-body systems due to the absence of a positive-definite probability measure \cite{Troyer:2004ge}. 
Thus, simulations of time evolution have been limited either to small numbers of particles that are amenable to exact diagonalization, or to one-dimensional gapped systems to which the time-dependent density matrix renormalization group \cite{White:1992zz,Schollwock:2005zz} can be applied. 
On the other hand, the growth of quantum entanglement limits the applicability of the latter for late times \cite{Cazalilla:2002,Vidal:2004,White:2004,Verstraete:2004}. 

\vskip -2pt
For practical purposes it is sufficient to consider the time evolution of the many-body system in terms of a quantum master equation for the reduced density matrix, where the degrees of freedom of the environment have been traced out. 
This logic follows the reality of experiments, where it is seldom possible to reconstruct the complete density matrix of the full coupled system.
When the coupling to the environment is weak and memory effects can be neglected, such an approach yields the Lindblad master equation \cite{Gorini:1976,Lindblad:1975ef,Breuer:2007}. 
In effect, this leads to a \textit{stochastic} quantum state evolution composed of two distinct parts: the \textit{continuous} nonunitary evolution with respect to an effective Hamiltonian $H_{\textrm{eff}} = H - \frac{i}{2} \sum_{\alpha} \gamma_{\alpha} L^{\alpha}{}^{\dagger} L^{\alpha}$ and the application of a set of \textit{discrete} quantum jump operators $L^{\alpha}$ \cite{Breuer:2007}. 
Generically, Hermitian jump operators result in the inevitable heating of the system that ultimately leads to an infinite-temperature ensemble, regardless of the initial state.
Much more interesting are non-Hermitian jump operators, which can be engineered in order to prepare specific input states for quantum simulation \cite{Diehl:2008b,Verstraete:2009,Bardyn:2013,Budich:2015}, e.g., using trapped ions \cite{Poyatos:1996,Barreiro:2011,Schindler:2013} or ultracold atoms in optical lattices \cite{Weimer:2010}. 
An intriguing proposal put forward in this context is a mechanism for ``dissipative cooling'' into a Bose-Einstein condensate (BEC) \cite{Diehl:2008,Kraus:2008,Schindler:2013}. 
Similar considerations based on dissipative quantum dynamics also play an important role in quantum information processing \cite{Raussendorf:2001,Childs:2002,Nielsen:2003,Aliferis:2004,Diehl:2011,Hu:2015} or entanglement generation \cite{Krauter:2011}. 
Essentially, for all of the considered cases, the system eventually reaches a nonequilibrium ensemble that is usually known exactly by construction. 
However, the intermediate real-time evolution of a macroscopic quantum system starting from a given initial state has so far remained largely out of reach.

In this Rapid Communication, we study the purely dissipative dynamics of a strongly correlated quantum spin system on a hypercubic $d$-dimensional lattice.
We derive the equations of motion for local observables and show that the resulting system of linear equations can be solved efficiently. 
This allows us to investigate the real-time dynamics of Bose-Einstein condensation via dissipation. 
We study the dependence of the dissipative gap on the system size and find nontrivial scaling behavior. 
We demonstrate that this has interesting implications for the mechanism of nonequilibrium condensation.

{\it Dissipative $s=1/2$ quantum spin model.} 
In the following, we consider a purely dissipative process ($H = 0$) for quantum spins at zero temperature. 
The spin operators, $s_x^a \equiv \tfrac{1}{2} \sigma_x^a$, $a = 1,2,3$, and $s_x^\pm = s_x^1\pm i s_x^2$, are defined in terms of Pauli matrices on each of the $N = L^d$ sites of a regular periodic lattice.
The real-time evolution of the reduced density matrix $\rho$ is assumed to be governed by a quantum master equation in the Lindblad form \cite{Gorini:1976,Lindblad:1975ef} that is characterized by a single dissipative process,
\eq{\label{eq:lindblad}
  \frac{d}{dt} \rho = \boldsymbol{\mathcal{L}} \rho \equiv \gamma \sum_{\langle x, y\rangle} \left( L_{x y}\rho \hsp L_{x y}^\dagger-\frac{1}{2} \big\{L_{x y}^\dagger L_{x y},\rho\big\} \right) \ .}
\vskip -3pt
The Lindbladian $\boldsymbol{\mathcal{L}}$ is defined in terms of non-Hermitian operators $L_{xy}=\frac{1}{2}(s_x^++s_y^+)(s_x^--s_y^-)$ that act on adjacent lattice sites $\langle x,y\rangle$. 
They map any two-particle spin-singlet state to the spin triplet, while conserving the total spin projection $S^3 = \sum_x s_x^3$ along the quantization axis, and annihilate the spin triplet. 
Here, $\gamma$ is the rate that we assign to the process. 
Starting from an arbitrary initial state, Eq.\ \eqref{eq:lindblad} eventually drives the system into a totally symmetric global superposition state \cite{Diehl:2008,Kraus:2008,Schindler:2013}.
Note that the \mbox{$s = 1/2$} quantum spin model can be mapped to a system of hard-core bosons \cite{Matsubara:1956}, where the spin operators $s_x^+$, $s_x^-$, and $s_x^{3}$ on each of the lattice sites are identified with the bosonic creation and annihilation operators $b_x^{\dagger}$, $b_x$, and $ b_x^{\dagger} b_x - 1/2$, respectively. 
Thus, by virtue of this mapping, the same dissipative process \eqref{eq:lindblad} can be viewed as symmetric delocalization of hard-core bosons over adjacent sites, with a BEC of hard-core bosons as the resulting final dark state.

\begin{figure}[!t]
\includegraphics[width=0.45\textwidth]{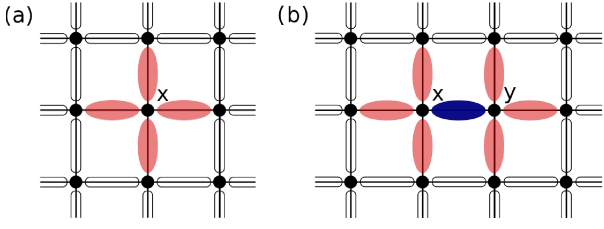}
\caption{The jump operators $L_{xy}$ act on all $dL^d$ pairs of nearest-neighbor sites on the regular periodic lattice with coordination number $n_c = 2 d$. This is illustrated above for the example of the two-dimensional lattice: (a) For each local operator, as, e.g., the local spin $s_{x}^a$, we attach $n_c$ jump operators (red) that contribute according to Eq.\ \eqref{eq:op_1}. (b) For two-point functions, as, e.g., $C_{xy}$, we need to consider two cases: If $x$ and $y$ are nearest neighbors, we attach one jump operator that connects both sites $x$ and $y$ (blue) [cf.\ Eq.\ \eqref{eq:tp_1}], while $n_c-1$ jump operators are assigned separately to $x$ and $y$ ({\it red}) [cf.\ \mbox{Eq.\ \eqref{eq:tp_2}}]. On the other hand, if $x$ and $y$ are nonadjacent, we attach $n_c$ jump operators to both $x$ and $y$.}
\label{fig1:lindblads}
\end{figure}

The nonvanishing eigenvalues of the Lindbladian $\boldsymbol{\mathcal{L}}$ with a negative real part $\operatorname{Re} \lambda_i < 0$ determine the relaxational modes that govern the real-time dynamics of the system. 
In particular, the mode corresponding to the eigenvalue with the largest real part,
\eq{\label{eq:gap}
 \Delta \equiv -\max_{i}\operatorname{Re}{ \lambda_i } > 0 \ ,}
dominates the asymptotic approach towards the nonequilibrium steady state. 
For the purely dissipative process Eq.\ \eqref{eq:lindblad}, the only scale in the problem is provided by the system size $N$ (i.e., the total number of particles). 
In the following, we show that even in the absence of a Hamiltonian, Eq.\ \eqref{eq:lindblad} leads to a nontrivial finite-size scaling of the dissipative gap for sufficiently large $N$. 
Note that the finite-size scaling of $\Delta$ has been studied in detail for various bosonic and fermionic systems in $d = 1$ dimension and Hermitian jump operators \cite{Esposito:2005,Prosen:2008,Eisler:2011,Znidaric:2011,Horstmann:2013,Cai:2013,Znidaric:2015}.

For any observable $O = O[s]$, the equation of motion for its expectation value $\mathcal{O}(t) =\operatorname{tr} \rho(t)\hsp O $ is given by
\eq{\label{eq:adlindblad}
 \frac{d}{dt} \mathcal{O} = \frac{\gamma}{2} \sum_{\langle x,y \rangle}\operatorname{tr} \rho \left\{ L_{xy}^\dagger\big[O,L_{xy}\big]+\big[L_{xy}^\dagger,O\big]L_{xy}\right\} \ .}
In general, the commutator terms in \eqref{eq:adlindblad} will induce new operators.
That is, the equations of motion for $m$-point functions typically depend on $(m+1)$-point functions --- the dynamical equations form an infinite hierarchy which cannot be solved in closed form. 
Notably, for Hermitian jump operators, conditions can be derived under which the hierarchy closes \cite{Zunkovic:2014}.
It seems that no definite statements have been made so far that establish whether a given non-Hermitian Lindblad process will lead to a closed system of equations. 
Here, we provide an explicit example where the hierarchy also closes, namely, for the non-Hermitian jump operators $L_{xy}$.
This allows us to study the finite-size scaling of $\Delta$ as well as the real-time dynamics of the dissipative process in arbitrary dimensions from first principles.

{\it One- and two-point correlation functions.} It is instructive to consider first the evolution of the spin components $\mathpzc{s}_x^a (t) = \operatorname{tr} \rho(t)\hsp s_x^a$. 
Note that the commutator terms in Eq.\ \eqref{eq:adlindblad} contribute only if the jump operators $L_{xy}$ are attached locally to the spin operator $s_x^a$ [cf.\ Fig.\ \ref{fig1:lindblads}(a)]. 
We determine
\eq{\label{eq:op_1}
  L^\dagger_{xy}\big[s_x^a,L_{xy}\big]+\big[L^\dagger_{xy},s_x^a\big]L_{xy}=\frac{1}{2}\big(s_y^a-s_x^a\big) \ ,}
and obtain the following diffusion equation for the local magnetization,
\eq{\partial_{t} \mathpzc{s}_x^a = (\gamma/4)\hsp \Delta_x \mathpzc{s}_x^a \ ,}
where $\Delta_xf_{x} \equiv \sum_{\mu =1}^{d} \left(f_{x-\hat{\mu}}-2f_x+f_{x+\hat{\mu}}\right)$ corresponds to the discretized Laplacian, and $\hat{\mu}$ denotes the unit vector in the $\mu$ direction on the spatial lattice.

Nontrivial correlations in the system are encoded in nonlocal operators such as $C_{xy}=s_x^+s_y^-+s_x^-s_y^+$ and $D_{xy}=s_x^3 s_y^3$. 
Note that $C_{xx}=4D_{xx}=\mathbbm{1}$, while the zero-momentum component of the Fourier-transformed two-point function $\mathcal{C}_{xy} (t) = \operatorname{tr} \rho(t)\hsp C_{xy}$ corresponds to the condensate fraction. Note, that in the following we use calligraphic fonts to denote ensemble averaged quantities.
Evaluating the commutators on the right-hand side of Eq.\ \eqref{eq:adlindblad} for the operator $C_{xy}$, we obtain
\eqa{\label{eq:tp_1}
  L^\dagger_{xy}\big[C_{xy},L_{xy}\big]+\big[L^\dagger_{xy},C_{xy}\big]L_{xy} &=& \mathbbm{1}-2C_{xy}-4D_{xy} \, , \\
  \label{eq:tp_2}
  L^\dagger_{xy}\big[C_{xz},L_{xy}\big]+\big[L^\dagger_{xy},C_{xz}\big]L_{xy} &=& \frac{1}{2}\big(C_{yz}-C_{xz}\big) \, ,}
where $x$ and $y$ correspond to adjacent sites. 
We point out that the operator $D_{xy}$ is generated \textit{only} if $x$ and $y$ are nearest-neighbor sites [cf.\ Fig.\ \ref{fig1:lindblads}(b)]. 
Considering the commutator terms for $D_{xy}$, we get
\eqa{L^\dagger_{xy}\big[D_{xy},L_{xy}\big]+\big[L^\dagger_{xy},D_{xy}\big]L_{xy} &=& 0 \, , \\
  L^\dagger_{xy}\big[D_{xz},L_{xy}\big]+\big[L^\dagger_{xy},D_{xz}\big]L_{xy} &=& \frac{1}{2}\left(D_{yz}-D_{xz}\right) \, .}
While the diagonal contributions are constant in time, $\mathcal{C}_{xx}=4\mathcal{D}_{xx}=1$, we obtain the following linear system for the off-diagonal contributions,
\eqa{\label{eq:eom_1}
  \partial_{t} \mathcal{C}_{xy} &=& \frac{\gamma}{4}(\Delta_x+\Delta_y) \hsp \mathcal{C}_{xy} - \frac{\gamma}{2}\delta_{\langle x,y\rangle} \big(\mathcal{C}_{xy}+4\mathcal{D}_{xy}\big) \, , \\
  \partial_{t} \mathcal{D}_{xy} &=& \frac{\gamma}{4}(\Delta_x+\Delta_y) \hsp \mathcal{D}_{xy} - \frac{\gamma}{8}\delta_{\langle x,y\rangle} \big(1-4\mathcal{D}_{xy} \big) \, ,  \label{eq:eom_2}} 
where $\delta_{\langle x,y\rangle}$ is nonzero and equal to one only if $x$ and $y$ are adjacent sites. 
Given initial data for the two-point functions $\mathcal{C}_{xy} (t = 0)$ and $\mathcal{D}_{xy} (t = 0)$, the first-order system of equations $\partial_t \hsp(\mathcal{C}\hsp,\,\mathcal{D})^\top=\mathcal{M} \,(\mathcal{C}\hsp, \,\mathcal{D})^\top$ can be solved explicitly. 
The solutions are expressed in terms of a superposition of exponential functions, whose characteristic rates of decay are determined by the eigenvalues $\lambda_i$ of the linear differential operator $\mathcal{M}$. 
Using spatial translation invariance, we will characterize the real-time evolution in momentum space in terms of the Fourier modes $\mathcal{C}_p (t) = N^{-2}\sum_{x,y}{e^{i p_{\mu}(x-y)_{\mu}}\, \mathcal{C}_{xy} (t)}$, with $p_{\mu} = 2\pi n_{\mu}/L$ and $n_{\mu}\in\{0,\ldots,L-1\}$.

{\it Infinite-temperature initial ensemble.} 
First, we consider the time evolution starting from an incoherent thermal ensemble at infinite temperature, i.e.,
\eq{\rho(t = 0) = 2^{-N} \mathbbm{1} \ ,}
for which all off-diagonal entries of the correlation functions vanish: $\mathcal{C}_{xy}(t = 0)=\mathcal{D}_{xy}(t = 0)=0$, $x \neq y$. 
The choice of initial conditions and subsequent dynamics can be likened to the following scenario: 
The system is initially prepared in the infinite-temperature ensemble and afterwards quenched to zero temperature, where the system is finally driven by the continuous application of quantum jump operators $L_{xy}$. 
While the diagonal elements $\mathcal{C}_{xx}(t)=4\mathcal{D}_{xx}(t)=1$ as well as the off-diagonal elements $\mathcal{D}_{xy}(t)=0$ remain constant as the system evolves in time, we observe that nontrivial off-diagonal correlations are generated for $\mathcal{C}_{xy}(t)$. 
This is clear, since by construction Eq.\ \eqref{eq:lindblad} leads to the following ensemble,
\eq{\label{eq:inftemp_final}\rho(t \rightarrow \infty) = 2^{-N} \sum_{n = 0}^N {N \choose n} |D(N,n) \rangle \langle D(N,n) | \ ,}
where $|D(N,n) \rangle \equiv |N/2 , -N/2 + n \rangle$ corresponds to the totally symmetric Dicke state, with $\vec{S}^2 |D(N,n)\rangle = N ( N + 2 ) / 4 |D(N,n) \rangle$ and $S^3 |D(N,n) \rangle = (-N/2 + n ) |D(N,n) \rangle$. 
Asymptotically, the ensemble \eqref{eq:inftemp_final} is characterized by
\eq{\mathcal{C}_p^{\infty}\equiv\lim_{t\to\infty}\mathcal{C}_{p}(t)= 1/2 \hsp\delta_{p,0} + 1/(2 N) \ .}

\begin{figure}[!t]
\centering
\includegraphics[width=0.45\textwidth]{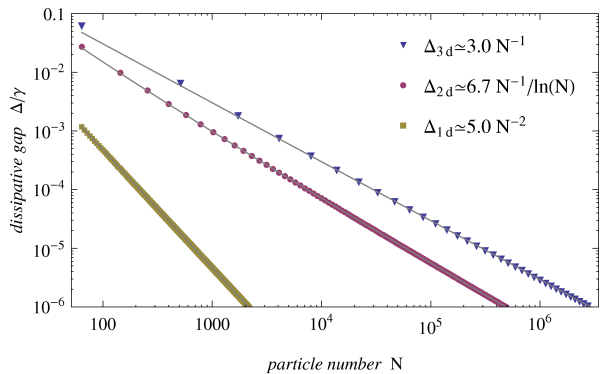}
\caption{Dissipative gap $\Delta$ in units of $\gamma$ as a function of system size (total particle number) $N$ on a double-logarithmic scale for dimensions $d=1$ ({\it squares}), $d=2$ ({\it dots}), and $d=3$ ({\it triangles}).}
\label{fig2:gap}
\end{figure}

\begin{figure}[!t]
\centering
\includegraphics[width=0.45\textwidth]{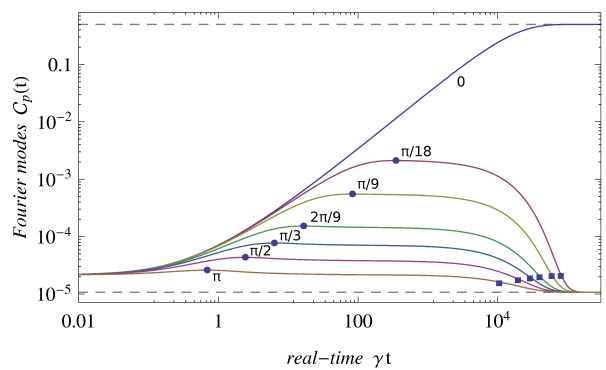}
\caption{Real-time evolution of a system in $d=3$ dimensions consisting of $N = 36^3$ particles initialized in the infinite-temperature thermal ensemble. 
Selected Fourier modes $\mathcal{C}_p(t)$, with $p_{\mu} = 2 \pi n \hsp\delta_{\mu,1}/ L$, $n = 1 , \ldots, L/2$, are shown, illustrating the growth and subsequent decay of correlations. 
The dashed lines indicate the asymptotic values $\mathcal{C}_p^{\infty}$; different points mark the two characteristic time scales $t_1\sim |\!| p |\!|^{-2}$ (\textit{dots}) and $t_2\sim |\!| p |\!|^{-1/3}$ (\textit{squares}).}
\label{fig3:evolution}
\end{figure}

\begin{figure}[!t]
\centering
\includegraphics[width=0.48\textwidth]{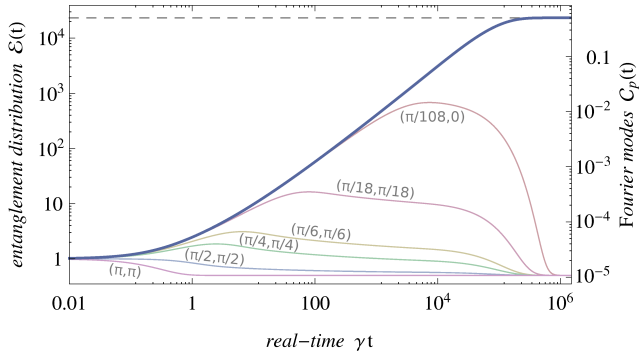}
\caption{Real-time evolution in $d=2$ dimensions for a system consisting of $N = 216^2$ particles prepared in an incoherent initial ensemble with total spin $S^3 = 0$. 
 We show the time-dependent cumulative entanglement distribution $\mathcal{E}(t)$ (\textit{blue}) and compare it to the behavior of the Fourier modes $\mathcal{C}_p(t)$, with $p_{\mu} = 2 \pi n_{\mu} / L$, $n_{\mu} = 1 , \ldots, L/2$ (\textit{labeled}). The dashed line indicates the asymptotic value for the entanglement distribution $\mathcal{E}^{\infty} = N$.}
\label{fig4:evolution}
\end{figure}

We solve the equations of motion Eqs.\ \eqref{eq:eom_1} and \eqref{eq:eom_2} via numerical diagonalization of the linear operator $\mathcal{M}$.
We observe that the dissipative gap $\Delta$ that governs the asymptotic behavior exhibits a nontrivial finite-size scaling that strongly depends on the dimension $d$, (cf.\ Fig.\ \ref{fig2:gap}); for large $N$, we find
\eqa{\label{eq:dimensionality_1}
 \Delta_{1d}^{-1}&\sim& N^2 \, , \\ \label{eq:dimensionality_2}
 \Delta_{2d}^{-1}&\sim& N \ln N \ , \\ \label{eq:dimensionality_3}
 \Delta_{3d}^{-1}&\sim& N \, .}
\vskip 5pt
An intriguing consequence of this behavior is that the dissipative process Eq.\ \eqref{eq:lindblad} increases in efficiency with dimension $d$. 
That is, for fixed particle number $N$, the asymptotic regime is reached earlier in time. The real-time evolution of some selected Fourier modes is shown in Figs.\ \ref{fig3:evolution} and \ref{fig4:evolution}. 
We identify three dynamic regimes for the low-lying momentum modes.

The \textit{regime of initial growth} is characterized by the generation of correlations due to the quasilocal action of the Lindblad operators, which gradually correlate quantum spins over ever larger distances.
Each of the eigenmodes with eigenvalues $\lambda_i$ contributes to the time dependence of $\mathcal{C}_{xy}$. 
The time $t_1$, when the low-lying modes $(p\neq 0)$ reach their maximum value, scales with inverse momentum squared, $t_1\sim |\!| p |\!|^{-2}$. 
Of course, this behavior is tied to the Laplacian Eqs.\ \eqref{eq:eom_1} and \eqref{eq:eom_2}, which also explains the quadratic scaling $t_1\sim L^2$ with the linear lattice extent, independent of dimension.

The lowest-lying eigenmode with a characteristic rate of decay $\Delta$ starts to dominate the dynamics in the subsequent \textit{transient regime}. 
All low-lying Fourier modes are seen to decay exponentially, $\mathcal{C}_{p\neq 0}\sim\exp(-\Delta t)$, until the influence of the asymptotic value $\mathcal{C}_{p = 0}^{\infty}$ becomes relevant. 
Thus, the dynamics of the condensate completely determines the behavior of the higher momentum-modes. 
Owing to the nontrivial scaling behavior Eqs.\ \eqref{eq:dimensionality_1} -- \eqref{eq:dimensionality_3}, we observe a separation of scales between $t_1\sim L^2$ and $\Delta^{-1}$ for the Fourier modes $\mathcal{C}_{p\neq 0}$ \mbox{(cf.\ Fig.\ \ref{fig3:evolution})}.

The \textit{regime of asymptotic decay} is characterized by $\mathcal{C}_p\sim\mathcal{C}_p^{\infty}[1+\exp(-\Delta t)]$. 
The transition between the transient and asymptotic regime is most easily seen by examining the absolute value of $\partial_t\ln{\mathcal{C}_p}$: 
On a log-linear plot, the transient regime is clearly identified by horizontal lines with height $\sim \Delta$, whereas the leading asymptotic behavior corresponds to straight lines with negative slope $-\Delta$. 
We define the time scale $t_2$ in terms of the intersection point of the corresponding extrapolated curves and find that it scales nontrivially with momentum $t_2\sim |\!| p |\!|^{-1/3}$ for the low-lying momentum modes.

{\it BEC preparation and entanglement growth.} 
In the following, we assume that $N$ is even and that the system is initially prepared in an incoherent ensemble with $S^3 = 0$, i.e.,
\eqa{&&\rho(t = 0) = \!\!\!\! \sum_{1 \leq x_{1} < \ldots < x_{N / 2} \leq N} \!\!\!\! |\phi(x_1, \ldots , x_{N/2})\rangle \langle\phi(x_1, \ldots , x_{N/2})| \ , \nonumber\\[-10pt] && \label{eq:initial} \\[-8pt]
  &&\hspace{4pt} |\phi(x_1, \ldots , x_{N/2}) \rangle \equiv {N \choose N/2}^{-\frac{1}{2}} s_{x_1}^{+} s_{x_2}^{+} \cdots s_{x_{N/2}}^{+}  \, |\Omega\rangle \ ,}
where $|\Omega \rangle \equiv \left| \, \downarrow \downarrow \cdots \downarrow \, \right\rangle$. 
Starting from Eq.\ \eqref{eq:initial}, the quantum dissipative process \eqref{eq:lindblad} drives the system into the Dicke state,
\eq{\rho(t\rightarrow \infty) = |D(N,0) \rangle \langle D(N,0) | \ .}
Similar to the infinite-temperature initial ensemble, here, the time evolution is carried out at zero temperature where the Lindblad jump operators dominate the dynamics. 
To quantify the real-time dynamics of entanglement, we consider the moments of the total spin operator $S^a$. 
Given an ensemble with no genuine $M$-particle entanglement ($M < N$), we can derive an upper bound for these moments \cite{Stockton:2003,Toth:2007,Duan:2011}. 
The measured value for $\mathcal{C}_{p = 0} = 2 \operatorname{tr} \rho\hsp [\vec{S}^2 - (S^3)^2 ]$ is used to check whether this bound is violated. 
If this is indeed the case, then we have shown that the time-evolved ensemble has $M$-qubit entanglement. 
Note that the Lindblad process conserves $S^{3}$ and therefore $\mathcal{D}_{p = 0}(t) = 0$ for all $t\in [0, \infty)$ and $\operatorname{tr} \rho\hsp (S^{3})^2 = \left(\operatorname{tr} \rho\hsp S^{3}\right)^2$.

To verify genuine $M$-particle entanglement at any given time, we need to check whether the following inequality is satisfied: $\mathcal{C}_{p = 0} > (M + 1) / ( 2 N )$ \cite{Duan:2011}. 
We use this relation to define the \emph{cumulative entanglement distribution}:
\eq{\mathcal{E}(t) = \max \left( 1 , \lceil 2 N \mathcal{C}_{p = 0}(t) - 2 \rceil \right) \ .}
It serves as a measure for the total number of entangled qubits in a given ensemble $\rho$. 
Note that its time dependence is fully determined by the condensate fraction $\mathcal{C}_{p = 0}$. 
At initial time $t = 0$, in the fully mixed state, $\mathcal{C}_{p = 0}(t = 0) = 1/N$, and we observe that there is no entanglement, $\mathcal{E}(t = 0) = 1$. 
In the infinite time limit, we obtain $\mathcal{C}_{p = 0}^{\infty} = 1/2 + 1/N$, and therefore $\mathcal{E}^{\infty} = N$. 
Note that well before the asymptotic regime $t\rightarrow \infty$ is reached, at time $t\sim \Delta^{-1}$, near to $N$ particles are mutually entangled: $\mathcal{E}(t) \simeq N \left[1 - \exp( - \Delta t)\right]$.
This clearly demonstrates the efficacy of the purely dissipative process Eq.\ \eqref{eq:lindblad} for the purpose of entanglement generation. The preparation of a macroscopic BEC in a two-dimensional lattice and the time evolution of the corresponding cumulative entanglement distribution is illustrated in Fig.\ \ref{fig4:evolution}.
Note that we have chosen the same value of the total particle number $N$ in both Figs.\ \ref{fig3:evolution} and \ref{fig4:evolution} to show that, for fixed $\gamma$, it takes longer for the condensate $\mathcal{C}_{p = 0}$ to reach its asymptotic value $\mathcal{C}_{p = 0}^{\infty}$ in $d = 2$ than in $d = 3$ dimensions. Currently, the numerical determination of the time evolution for the low-lying modes allows us to solve for system sizes of up to $N \approx 80^3$ particles.

{\it Conclusions.} In this work, we have investigated the real-time dynamics of a purely dissipative $s = 1/2$ quantum spin system. 
This serves as an interesting model to study the application of non-Hermitian jump operators for state preparation and entanglement generation. The same dissipative process with competing unitary dynamics has been considered to some extent in previous work, e.g., in Ref.\ \cite{Schindler:2013} for a system of up to $N = 10$ particles in the framework of a discrete time evolution generated by a Kraus map, while a mean-field approach was used to study linearized theories around a weakly perturbed dark state \cite{Diehl:2008}. However, here we are able to resolve the complete time evolution for a \textit{macroscopic} number of particles, albeit in the absence of a competing Hamiltonian dynamics.
The large but finite system size provides a scale that determines the characteristic time for the evolution of correlations.
This allows us to extract the asymptotic scaling of the dissipative gap.
In particular, we find a nontrivial finite-size scaling that depends on the dimension. 
The dissipative process becomes more efficient as the coordination number of the lattice is increased.
This certainly has interesting implications, e.g., for state preparation in ultracold atoms in optical lattices. 
Furthermore, we have shown explicitly how multiparticle entanglement is generated in real time and how the system evolves into a macroscopically entangled BEC.

So far, we have neglected the effect of a thermal bath as the system is driven at zero temperature. We plan to investigate the effect of thermal fluctuations on the real-time dynamics and the stability of the final dark state in a future publication. Another interesting question concerns the role of topological excitations for the dynamics which we plan to address.

We thank D.\ Banerjee, J.\ Berges, \mbox{H.\ P.\ B\"uchler}, S.\ Chandrasekharan, S.\ Diehl, E.\ Huffman, and P.\ Zoller for illuminating discussions. 
This research is funded by the European Research Council under the European Union's Seventh Framework Programme, FP7/2007-2013, 339220.

\end{document}